\documentclass[aps,prb,twocolumn,showpacs,superscriptaddress,10pt]{revtex4-1}\begin{tiny}\end{tiny}
\usepackage{longtable}
\usepackage{amsfonts}
\usepackage[dvipdfmx]{graphicx}
\usepackage{setspace}
\usepackage{rotating}
\usepackage{hyperref}
\usepackage[version=3]{mhchem}

\begin{document}

\title{Electronic and Nuclear Quantum Effects on the Ice XI/Ice Ih Phase Transition}

\author{Bet\"{u}l Pamuk}
 \affiliation{Physics and Astronomy Department, SUNY Stony Brook University, NY 11794-3800, USA}
 \affiliation{CNRS, UMR 7590 and Sorbonne Universit\'{e}s, UPMC Univ Paris 06, IMPMC - Institut de Min\'{e}ralogie, de Physique des Mat\'{e}riaux, et de Cosmochimie, 4 place Jussieu, F-75005, Paris, France}

\author{Philip B. Allen}
 \affiliation{Physics and Astronomy Department, SUNY Stony Brook University, NY 11794-3800, USA}
 
\author{M.-V. Fern\'andez-Serra}
 \email{maria.fernandez-serra@stonybrook.edu}
 \affiliation{Physics and Astronomy Department, SUNY Stony Brook University, NY 11794-3800, USA} 
 \affiliation{Institute for Advanced Computational Sciences, Stony Brook University, Stony Brook, New York 11794-3800, USA}

\date{\today}


\begin{abstract}

We study the isotope effect on the temperature of the proton order/disorder
phase transition between ice XI and ice Ih, using the quasiharmonic approximation combined with \textit{ab initio}
density functional theory calculations. 
We show that this method is accurate enough
to obtain a phase transition temperature difference between light ice (H$_2$O) and heavy ice (D$_2$O) of 6 K 
as compared to the experimental value of 4 K.
More importantly, we are able to explain the origin of the isotope effect on the much debated large temperature
difference observed in the phase transition. 
The source of the difference is directly linked to the physics behind the anomalous isotope effect
on the volume of hexagonal ice that was recently explained in [Phys. Rev. Lett. 108, 193003 (2012)].
These results indicate that the same physics might be behind the isotope effects in
transition temperatures between other ice phases.

\end{abstract}

\maketitle

\section{Introduction} 

The polymorphism of ice is revealed by its rich phase diagram \cite{Review2012,Petrenko}. 
The availability of different
proton configurations that satisfy Bernal-Fowler ``ice-rules" \cite{Bernal33} adds
another dimension to this phase diagram, given that the same crystalline
structure could exist in proton-ordered and disordered form.
This leads to additional phases, separated by
their corresponding order/disorder phase transitions,
as in the case of ice XI/ice Ih, ice IX/ice III, and ice VIII/ice VII \cite{Petrenko}.

In this paper we focus on the phase transition between proton-ordered (ice XI) and 
proton-disordered (ice Ih) hexagonal ice.
This phase transition has been subject of a large number of experimental\cite{Tajima82, Tajima1984, Matsuo1986, Wooldridge1988, Suga1997}
and theoretical studies\cite{Rick2005, Baez1995, Arbuckle2002, Vlot1999, Nada2003, Sanz2004, Sadlej1998, Kuo2005, Pan2008, Pan2010, Klein2005, Klein2006, Schonherr2014}. 
However, open questions remain about the mechanisms behind the phase transition
and the importance of nuclear quantum effects in the temperature of the transition \cite{SugaReview}.

Experimentally, it is difficult to observe the phase transition from ice Ih to ice XI.
A glass transition occurs at around 100-110 K \cite{Wooldridge1988, Suga1997},
diminishing the proton mobility and locking protons in their disordered positions,
before they orient to form the proton-ordered ice XI structure.
This is overcome by catalyzing ice Ih by KOH \cite{Tajima82, Tajima1984, Matsuo1986},
which allows the lattice parameters of both proton-ordered ice XI 
\cite{Leadbetter84, Howe1989, Jackson1995, Whitworth1996, Jackson1997, Whitworth1997} 
and disordered ice Ih \cite{Kuhs94, Kuhs12, Whitworth1996}
to be experimentally measured.
The order/disorder phase transition is achieved at 72 K for light \cite{Tajima82}
and 76 K for heavy ice \cite{Matsuo1986}.
Although the isotope effect on the phase transition temperature is measured to be 4 K, 
a theoretical explanation for this difference is still missing.
Ref. \onlinecite{Chan75, JohariJones76} associated the origin of the isotope effect on the
transition temperature to the difference in vibrational energies
between the two phases estimating a $\sim 27$ K isotope effect 
on the transition temperature,  
while Ref. \onlinecite{Matsuo1986} sought the explanation in the difference in 
reorientation of the dipole moments of heavy and light ices
and predicted a much smaller $\sim 1$ K isotope effect.
However, in both of these studies they assume no isotope effect on the volume
of the two ices and they also did not take into account the 
competing anharmonicities between the intra-molecular covalent bonds 
and the inter-molecular hydrogen bonds.
In this work, we reexamine the issue taking into account these two additional effects,
extending a previous work on the anomalies in the isotope effect on the
volume of ice \cite{Pamuk12}.

A large  literature is devoted to the study of the order/disorder
phase transition in hexagonal ice. Two main questions are discussed,
(i) the ordering nature in the low temperature phase, and (ii)
theory and simulation predictions for the phase transition temperature.
Experiments such as neutron diffraction\cite{Whitworth1996, Jackson1997, Whitworth1997},
as well as measurements performed under an electric field \cite{Jackson1995}
indicate that the ordered phase has ferroelectric order.
However, among theory and simulation works there is a large dispersion of results and lack of agreement
\cite{Rick2005, Baez1995, Arbuckle2002, Vlot1999, Nada2003, Sanz2004, Sadlej1998, Kuo2005, Pan2008, Pan2010, Klein2005, Klein2006, Schonherr2014}.
The predicted low T stable phase depends strongly 
on the choice of boundary conditions, electrostatic multipoles,
and treatment of long range interactions.
\cite{Rick2005, Baez1995, Arbuckle2002, Vlot1999, Nada2003, Sanz2004, Sadlej1998}.
In addition, semi-empirical force field  models fitted to reproduce experimental data are not accurate enough to
distinguish small energy differences 
between different proton orderings.

According to Ref. \onlinecite{Rick2005}, the TIP4P-FQ \cite{tip4p-fq} model predicts the proton-disordered phase 
to be stable, in agreement with Ref. \onlinecite{Sadlej1998} where other less known models were studied.
On the other hand popular water models like SPC/E \cite{SPC/E}, TIP4P \cite{tip4p}, TIP5P-E \cite{tip5p} 
and NvdE \cite{Nada2003} models predict the proton ordered phases to be more stable in the low temperature limit.
Among these studies, only the NvdE \cite{Nada2003} model predicts the ferroelectric-ordered phase as the lowest energy phase,
in agreement with experiments, while the other three models predict the stable phase to be antiferroelectric-ordered.
In this same study, it was also shown that modifying the polarizability of the KW-pol model 
was enough to favor ferroelectric ordering over disordered configurations at low T \cite{Sadlej1998}.
Therefore polarizability is an important factor in obtaining the correct potential energy surface.

The effect of proton disorder on the hexagonal ice structure has also been studied using
\textit{ab initio} density functional theory (DFT).
DFT calculations correctly reproduce the lattice structure \cite{Kuo2005},
and give the cohesive energy of ferroelectric-ordered ice to be larger than either antiferroelectric-ordered or disordered ices \cite{Pan2008, Pan2010}.
That is, ferroelectric-ordered ice is the stable ground state.
A recent DFT study of ice slabs shows that contrary to the bulk case, the antiferroelectric-ordered ice is more stable
in the case of thin films \cite{Parkkinen2014}.
DFT-based simulations have also been used to predict the phase transition temperature.
A Monte Carlo study, where DFT calculations of hydrogen bond configuration energies 
are used to parametrize a model to perform Monte Carlo simulations, 
predicted ice XI as the most stable phase, with a transition temperature of 98 K. \cite{Klein2005, Klein2006}
Another recent DFT-based Monte Carlo study of dielectric properties of ice,
predicted the temperature of the order-disorder phase transition to be 
around 70-80 K \cite{Schonherr2014}.
The advantage of DFT-based Monte Carlo simulations is that they can explore the configurational 
entropy of the free energy surface in good detail.
However, none of these calculations include zero point nuclear quantum effects,
or thereby investigate the transition temperature difference between different isotopes.
Our goal is to investigate the order/disorder
transition from ferroelectric-ordered ice or antiferroelectric-ordered ice
to disordered ice, with different isotopic compositions, from an \textit{ab initio}
perspective, including nuclear quantum effects.

In our study 
we will also compare a polarizable force field model, TTM3-F \cite{ttm3f}
to DFT calculations for the prediction of the most stable phase at low temperatures
including zero point corrections.

\section{Theory}

In a recent study, we explained the anomalous isotope effect
on the volume of ice \cite{Kuhs94, Kuhs12, Pamuk12},
by obtaining the free energy with \textit{ab initio} DFT
within the quasiharmonic approximation.
We have shown that the anticorrelation between  
the intra-molecular OH covalent bonds and the inter-molecular hydrogen bonds 
makes the volume per molecule of D$_2$O ice larger than that of H$_2$O ice \cite{Pamuk12}.

In this work, we extend our study of nuclear quantum effects to analyze the contribution to the
order/disorder phase transition using both \textit{ab initio} DFT functionals and the TTM3-F \cite{ttm3f} force field model.
We investigate both ferroelectric-ordered/disordered and antiferroelectric-ordered/disordered phase
transitions.
In addition, we analyze the importance of van der Waals forces,
by comparing a generalized gradient approximated functional, Perdew-Burke-Ernzerhof (PBE) \cite{PBE},
to a van der Waals functional\cite{DRSLL, SolerVdW}, vdW-DF$^{\rm{PBE}}$ .
We obtain the temperature dependence of the free energy for both ice phases using the quasiharmonic approximation (QHA),
and we compare the phase transition temperature of different isotopes.

\subsection{Free Energy within Quasiharmonic Approximation}

To account for nuclear quantum effects, quantum harmonic eigenstates are needed as a function
of volume $V$, at volumes near $V_0$, the ``frozen lattice'' zero pressure volume that minimizes
the Born-Oppenheimer energy, $E_0(V)$.  To lowest order in a Taylor series around $V_0$, we have
\begin{equation}
E_0(V)=E_0(V_0) + \frac{B_0}{2V_0} (V-V_0)^2
\label{eq:Bzero}
\end{equation}
and 
\begin{equation}
\omega_k(V) = \omega(V_0) \left( 1- \gamma_k \frac{V-V_0}{V_0} \right) .
\label{qh2}
\end{equation}
$B_0$ is the dominant part of the bulk modulus, omitting vibrational corrections which will be discussed
in a later paper.  The ``mode Gr\"uneisen parameters'' $\gamma_k$ are defined as
\begin{equation}
\gamma_k=-\frac{\partial(\ln \omega_k)}{\partial(\ln V)} =-\frac{V}{\omega_k} \frac{\partial\omega_k}{\partial V} .
\label{eq:grun}
\end{equation}
The phonon frequencies, $\omega_k$ are calculated at three different volumes.
The volume dependence of $\omega_k(V)$ is calculated to the linear order.
Then the Helmholtz free energy $F(V,T)$ ~\cite{Ziman}
of independent harmonic oscillators acquires a volume-dependence through $\omega_k(V)$,
\begin{eqnarray}
F(V,T) 
  &=& E_0(V) + \nonumber \\
  && \sum_k \left[ \frac{\hbar\omega_k(V) }{2} + 
            k_B T \ln \left(1-e^{-\hbar \omega_k(V) / k_B T}\right) \right] \nonumber \\
  && -TS_H
\label{eq:free}
\end{eqnarray}
The index $k$ runs over both phonon branches and phonon wave vectors within the Brillouin zone.
This ``quasiharmonic'' approximation is correct to first order for volume derivatives like 
$P=-(\partial F/\partial V)_T$.  Higher volume derivatives, such as $B(T)$, in general
may require higher volume derivatives of $E_0$ and $\omega_k$.  
As shown in our recent contributions \cite{Pamuk12, Herrero12}, the first derivative in eq. \ref{qh2} is a good 
approximation for hexagonal ice.  
The temperature dependence of volume $V_{F_{\rm min}}(T)$ is then found
in the usual way by minimizing $F(V,T)$ at fixed $T$, the same as setting $P(T)=0$.

The last part of the free energy, $S_H$, is the entropy of the proton disorder.
This term is zero for proton-ordered ice phases.
For the proton-disordered phase, ice Ih, we use the estimation by Pauling, $S_H=N k_B \ln(3/2)$, 
which was obtained by counting hydrogen orientations that obey the ice rules \cite{Pauling1935},
and experimentally confirmed for fully disordered cases \cite{Giauque1936, Giauque1933}.
We assume that this term does not change with temperature.

Lastly, the classical limit of the free energy is obtained by taking the high temperature limit of the QHA:
\begin{eqnarray}
F(V,T)&=&E_0(V)+ \nonumber \\ 
      &&\sum_k \left[ k_B T \ln \left(\frac{\hbar\omega_k(V(T))}{k_B T} \right) \right] - T S_H .
\label{eq:cla-free}
\end{eqnarray}

\subsection{Cohesive Energy}

To determine which structure is the most stable one at zero temperature,
cohesive energies of ices are calculated without ($E_c^0$) and with ($E_c$) zero point effects.
The cohesive energy is defined as the amount the energy of a molecule is lowered in a crystal relative to in vacuum:
\begin{equation}
E_c^0 = \frac{E_{0}^{{\rm ice}}}{N_{{\rm molecules}}} - E_{0}^{{\rm monomer}}
\label{eq:Ec0}
\end{equation}
\begin{equation}
E_c = \frac{F^{{\rm ice}}(V_{F_{\rm min}},0)}{N_{{\rm molecules}}} - E_{0}^{{\rm monomer}} - E_{vib}^{{\rm monomer}} ,
\label{eq:Ec}
\end{equation}
where the vibrational energy $\sum_k \hbar \omega_k /2$ of the three modes
of the monomer is $E_{vib}^{\rm{monomer}}$.
The classical cohesive energy, $E_c^0$ is defined in eq. \ref{eq:Ec0} using the Kohn-Sham energies of the ice and monomer;
and similarly, the quantum cohesive energy with zero point effects, $E_c$ is defined in eq. \ref{eq:Ec}.

\section{Simulation Details}
\subsection{System Description}

\begin{figure}[!ht]
        \centering
        \includegraphics[clip=true, trim=0mm 10mm 0mm 20mm, scale=0.18]{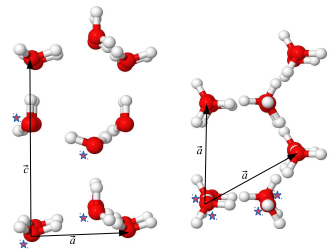} 
        \caption{Unit cell of the ferroelectric proton-ordered ice XI structure.
		The 4 molecules in the unit cell are labeled with a star symbol next to it, and $a$, and $c$ lattice vectors are shown.
		The image on the left is the side view of the x-z plane; the image on the right is the top view of the x-y plane.}
        \label{fig:iceXI}
\end{figure}

In order to predict the most stable phase of hexagonal ice in the zero temperature limit,
we performed total energy calculations of three hexagonal ices with different proton configurations.
(i) Ice XI. This is the ferroelectric-proton-ordered ice. 
Oxygen atoms are constrained to the hexagonal wurtzite lattice 
and hydrogen atoms are ordered such that ice XI has a net dipole moment along the $\hat{c}$ axis, 
shown in Fig. \ref{fig:iceXI}. 
Precise measurements of lattice structure of ice XI have shown ferroelectric ordering, with a net dipole moment along
the $\hat{c}$-axis. \cite{Leadbetter84, Howe1989, Jackson1995, Whitworth1996, Jackson1997, Whitworth1997}
There are 4 molecules per formula unit. However, TTM3-F calculations were performed for a 
$3a \times 2\sqrt{3}a \times 2c$ supercell with 96 molecules, with the same cell size as
disordered ice Ih.

\begin{figure}[!ht]
        \centering
        \includegraphics[clip=true, trim=0mm 35mm 0mm 45mm, scale=0.23]{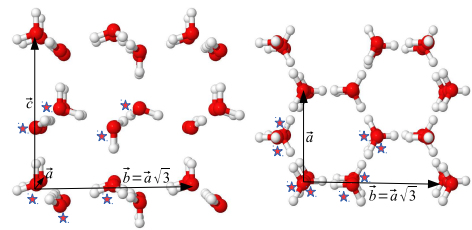}
        \caption{Antiferroelectric proton-ordered ice aXI structure.
		The 8 molecules in the unit cell are labeled with a star symbol next to it, and $a$, $b$, and $c$ lattice vectors are shown.
		The image on the left is the side view of the x-z plane; the image on the right is the top view of the x-y plane.}
        \label{fig:iceaXI}
\end{figure}

(ii) Ice aXI. We label the antiferroelectric-proton-ordered ice as ice aXI.
Ice aXI has 8 molecules in the unit cell.
The unit cell is doubled from the ferroelectric proton-ordered ice XI along the x-y plane,
with dipole moments of the neighboring molecules pointing in opposite directions
such that the system has no net dipole moment, as shown in Fig. \ref{fig:iceaXI}.
Similarly, we have used a unit cell of 8 molecules for the DFT, and 96 molecules for the TTM3-F calculations.

\begin{figure}[!ht]
        \centering
        \includegraphics[clip=true, trim=0mm 60mm 0mm 30mm, scale=0.23]{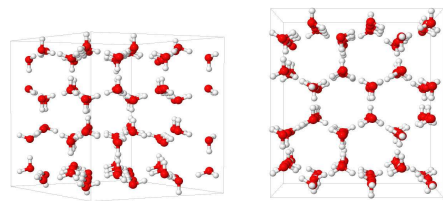} 
        \caption{Proton-disordered ice Ih structure. The image on the left is the side view of the x-z plane; the image on the right is the top view of the x-y plane.}
        \label{fig:iceIh}
\end{figure}

(iii) Ice Ih. 
Experimentally, the lattice structure of light(H$_2$O) and heavy(D$_2$O) 
proton-disordered hexagonal ice Ih
have been measured using both syncrotron radiation \cite{Kuhs94, Kuhs12} 
and neutron diffraction \cite{Whitworth1996}, with good agreement.
Oxygen atoms still have an underlying hexagonal lattice, 
while hydrogen atoms are disordered such that it has no net dipole moment. 
An example of this system is shown in Fig. \ref{fig:iceIh}.
To accommodate different proton-disordered configurations in ice Ih,
we have used a large cell of 96 molecules with dimension $3a \times 2\sqrt{3}a \times 2c$ in our DFT calculations.

We have computed five different 
96 molecule configurations of ice Ih using the TTM3-F model. They are generated with an algorithm 
that goes over all possible allowed proton configurations and produces structures with no net dipole moment \cite{Sadlej1998}.

\subsection{Simulation Procedure}

We used {\sc siesta} code \cite{SiestaPRBRC, SiestaJPCM} to perform DFT calculations within the generalized gradient 
approximation (GGA) to the exchange and correlation (XC) functional. 
The calculations use PBE and vdW-DF$^{\rm{PBE}}$ functionals, \cite{PBE, DRSLL, SolerVdW}
to compare non-local van der Waals effects with semi-local GGA approximations.
These density functionals have previously been shown to 
give good results for volume calculations of hexagonal ice Ih. \cite{Pamuk12}

Full structural relaxations for calculating the $E_0(V)$ curve are performed with t$\zeta$+p basis.
For these relaxations, we have used a real-space mesh cut-off of 500 Ry 
for the integrals, electronic k-grid cut-off of 10 \AA~ (corresponding to 38 k-points) for unit cell 
calculations of ice Ih, force tolerance of 0.001 eV/\AA~, and a density matrix tolerance of $10^{-5}$ electrons. 
Instead of doing a variable cell optimization, we calculate the energy 
of the relaxed structure at a fixed volume for each lattice parameter.

Even though results with the t$\zeta$+p basis are accurate enough to obtain general structural properties \cite{Pamuk12},
for precise order-disorder free energy values, the energy must be very well converged.
Recently, a systematic method to obtain the finite-range atomic basis sets for liquid water and ice has been proposed \cite{Corsetti2013}.
We use the quadruple-$\zeta$ double polarized (q$\zeta$+dp) basis obtained with the new proposed framework.
We calculate the energy of the structures again, with the q$\zeta$+dp basis, 
without relaxation.

For the t$\zeta$+p basis, the error compared to the q$\zeta$+dp basis is $-0.23\%$ in lattice constant $a$, $-0.28\%$ in $c$,
and $-0.71\%$ in the total volume.
The change in the energy, $E_0(V_0)$ from t$\zeta$+p basis to the q$\zeta$+dp basis without relaxation is 948.6 meV.
Further relaxing the structures with the q$\zeta$+dp basis does not change the lattice parameters,
and changes the energy only by 1.3 meV.
Details of this calculation and the lattice parameters with q$\zeta$+dp basis are given in the Supplemental Material (SM) \cite{SM} .

For the free energy calculations which include the nuclear quantum effects, 
the vibrational modes are calculated using the frozen phonon approximation.
All the force constant calculations are performed with the t$\zeta$+p basis.
There are two reasons for this:
the t$\zeta$+p basis gives a good first approximation to the configurational information,
and the q$\zeta$+dp basis is costly in computer time.
In addition, the largest error in the free energy calculations comes from the initial $E_0(V)$ contribution,
which we reduce significantly as explained above.
The error in the zero point energy contribution is much smaller than the electronic energy error.
The force constant calculations of proton-ordered ice XI structure use a finer real-space mesh cut-off of 800 Ry 
and an atomic displacement of $\Delta x = 0.06$ \AA . 
Similarly, the force constant calculations of proton-disordered ice Ih use a real-space mesh cut-off of 500 Ry 
and an atomic displacement of $\Delta x = 0.08$ \AA~ for the frozen phonon calculations.
The acoustic sum rule has been used throughout the study.

The phonon frequencies, $\omega_k(V_0)$ and Gr\"uneisen parameters $\gamma_k(V_0)$ are obtained by diagonalizing 
the dynamical matrix, computed by finite differences from the atomic
forces in a $(3 \times 3 \times 3)$ supercell, at volumes slightly below and above $V_0$.
We tested these parameters to obtain force constants in phonon calculations, 
so that the Gr\"uneisen parameter calculations have minimum noise \cite{Pamuk12}. 
The Gr\"uneisen parameters are calculated for 3 volumes
corresponding to isotropic expansion and compression around the minimum.
In order to cover the full Brillouin zone of ice XI and ice aXI, 729 k-points are selected, 
dividing each reciprocal lattice vector into 9 equal sections.

\begin{table*} [hbt!] \footnotesize
\caption{Classical (E$_c^0$) and quantum ($E_c$ ) cohesive energies, in meV. Quantum values include zero point effects.}
\begin{center}
		\begin{tabular}{l l c c c c}
		\hline
		\hline
		\cline{1-6}
		FF/XC                  & Ice & E$_c^0$ & H$_2$O & D$_2$O & H$_2\text{}^{18}$O \\ 
		\hline
		TTM3-F                 & Ih  & 601.07$\pm$0.22 & 521.17$\pm$0.22 & 536.24$\pm$0.22 & 522.87$\pm$0.23 \\
		TTM3-F                 & aXI & 600.30  & 520.33 & 535.41 & 522.04 \\
		TTM3-F                 & XI  & 599.71  & 520.11 & 535.11 & 521.81 \\
		PBE                    & Ih  & 620.44  & 502.07 & 526.70 & 504.15 \\
		PBE                    & aXI & 626.21  & 507.17 & 531.95 & 509.25 \\
		PBE                    & XI  & 629.06  & 509.02 & 534.07 & 511.11 \\	
		vdW-DF$^{\rm{PBE}}$    & Ih  & 723.94  & 601.64 & 627.65 & 603.64 \\
		vdW-DF$^{\rm{PBE}}$    & aXI & 725.53  & 602.58 & 628.80 & 604.57 \\
		vdW-DF$^{\rm{PBE}}$    & XI  & 728.75  & 605.18 & 631.52 & 607.18 \\
		\hline
		\hline
		\end{tabular}
\end{center}
\label{table:Ec}       
\end{table*}

\section{Results}

\subsection{Order-Disorder Phase Transition}

To understand the phase transition between proton-disordered  and
proton-ordered ice, we calculated the cohesive energy from eqs. \ref{eq:Ec0}, \ref{eq:Ec}.
The cohesive energy including the zero-point nuclear quantum effects are 
also presented in Table \ref{table:Ec}.
The results from different proton-disordered configurations using the TTM3-F model all lie within $\pm 0.22$ meV of each other,
as indicated in the first line of Table \ref{table:Ec}.
The change in the cohesive energy due to the residual entropy of hydrogen disorder is on the order of 0.22 meV, 
which means that our quantitative prediction of the most stable phase is within this range.

Both DFT functionals predict stability to decrease in the order ice XI $\rightarrow$ ice aXI $\rightarrow$ ice Ih.
This agrees with the experiments that the structure of the ordered phase is ferroelectric.
On the other hand, TTM3-F predicts the stability order to be the reverse, ice Ih $\rightarrow$ ice aXI $\rightarrow$ ice XI,
giving the wrong ground state and no phase transition.
Furthermore, considering the error in the cohesive energy, it is impossible to predict the correct stable
phase at the zero temperature limit with this model.
The phase transition can only be obtained with the DFT calculations.

In order to analyze the proton order-to-disorder phase transition temperature, 
we study the Helmholtz Free energy at zero pressure.
We evaluate the volume dependence of free energy, F(V) at fixed temperature
and find the value of free energy minimum, $F(V_{F_{\rm min}}(T))$.
Therefore, we obtain a temperature dependence of free energy, by evaluating free energy minimum 
for each temperature, $F(V_{F_{\rm min}}(T))$, using eq. \ref{eq:cla-free} or \ref{eq:free}.

In the classical limit of the free energy, 
without considering nuclear quantum effects, as given in eq. \ref{eq:cla-free}, 
DFT predicts a phase transition, regardless of the chosen functional.
In addition, both semi-local PBE and non-local vdW-DF$^{\rm{PBE}}$ functionals overestimate the phase
transition temperature, when nuclear quantum effects are not included in the calculations.
On the other hand, the TTM3-F force field model does not correctly predict the stable phase
in the low temperature limit, and the difference between the free energies of the two phases
increases with temperature. Therefore, it does not show a phase transition.
We present the full temperature dependence of classical free energy in the SM \cite{SM}.
To understand how each component of eq. \ref{eq:cla-free} contributes to the total free energy,
we also present the temperature dependence of $E_0(V_{F_{\rm min}}(T))$, and $-TS$ terms separately in the SM \cite{SM}.

\begin{figure}[!ht]
        \centering
                \includegraphics[clip=true, trim=0mm 0mm -5mm -12mm, scale=0.5]{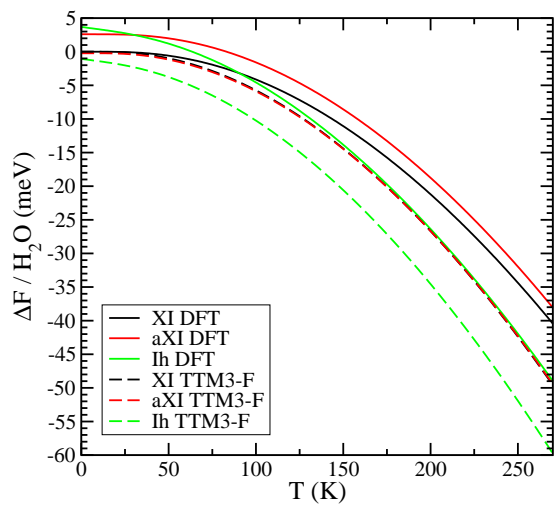} 
        \caption{Relative free energy per molecule including the quantum zero point effects as a function of temperature.
        The lines show DFT results with vdW-DF$^{\rm{PBE}}$ functional and the dashed lines are the results with the TTM3-F model.
        DFT correctly predicts the most stable phase as ice XI for low temperatures, with an energy difference of $\sim$ 4.5 meV.
	For low temperatures, the TTM3-F model predicts ice Ih as the stable phase with $\sim$ 1 meV energy difference at zero temperature;
        and a separation of energy at higher temperatures, making the prediction correct for high temperatures only.
	The results of the TTM3-F model for ice XI and ice aXI represented by black and red dashed lines respectively
	are almost indistinguishable in this scale.}
        \label{fig:F_T}
\end{figure}

\subsection{Isotope Effects in the Transition Temperature}

Going further, the nuclear zero point effects are calculated to compare the predicted phase
transition temperature to the experiments.
Fig. \ref{fig:F_T} shows the temperature dependence of the free energy with zero point effects for H$_2$O.
At all temperatures, TTM3-F model predicts ice Ih to be the stable phase.
DFT correctly predicts the most stable phase as the ferroelectric-ordered ice XI for low temperatures, with an energy difference of $\sim$ 4.5 meV.
As the temperature increases, there is a crossing at $T=91$ K and the proton-disordered ice Ih becomes the stable phase beyond this temperature for H$_2$O.
DFT predicts antiferroelectric ice aXI to have lower free energy than ice Ih at low T, but at all T, ferroelectric ice XI is preferred to ice aXI.
The crossover from the antiferroelectric-ordered ice aXI to 
proton-disordered ice Ih is at much lower temperatures,
because ice aXI is less cohesive than the ferroelectric-ordered ice XI.
Therefore, we establish that with DFT, the most stable phase at low temperatures is the ferroelectric-ordered ice XI, 
in agreement with the experiments.
For the rest of the transition temperature discussion, we will focus on the ferroelectric-ordered to disordered transition, ice XI/ice Ih.

\begin{table*} [!t] \footnotesize
\caption{The classical (T$_c^0$) and quantum (T$_c$) proton order to disorder transition temperature, $T_c$ (K) including zero point effects
        for ice Ih-ice XI and ice Ih-ice aXI.
        The ratio of the temperature for different isotopes is given as R(D)=D$_2$O/H$_2$O and R($^{18}$O)=H$_2\text{}^{18}$O/H$_2$O,
	and the isotope effect on the temperature with respect to the H$_2$O transition temperature is also given as
        the isotope effect percentage: IS(A-B)=$\frac{T(A)}{T(B)}-1$.}
\centering
                \begin{tabular}{l l r r r r r r r r} 
                \hline
                \hline
                \cline{1-10}
                Ice & Method                           & T$_c^0$ & H$_2$O & D$_2$O & H$_2\text{}^{18}$O & R(D) & R($^{18}$O) & IS(D-H) & IS($\text{}^{18}$O -$\text{}^{16}$O) \\
                \hline
                aXI & PBE                              & 153     & 151    & 156    & 151 & 1.03 & 1.00 & $ +3.31\%$ & $ 0.00\%$ \\
                aXI & vdW-DF$^{\rm{PBE}}$              &  42     & 30     & 35     & 30  & 1.17 & 1.00 & $+16.67\%$ & $ 0.00\%$ \\
                XI  & PBE                              & 221     & 202    & 215    & 203 & 1.06 & 1.01 & $ +6.44\%$ & $+0.50\%$ \\
                XI  & vdW-DF$^{\rm{PBE}}$              & 105     & 91     & 97     & 90  & 1.07 & 0.99 & $ +6.59\%$ & $-1.10\%$ \\
                XI  & Expt \cite{Tajima82, Matsuo1986} &         & 72     & 76     &     & 1.06 &      & $ +5.56\%$ &  \\
                \hline
                \hline
                \end{tabular}
\label{table:Tc}
\end{table*}

Inclusion of zero point effects also allows us to obtain the isotope effect in the phase transition temperature,
since it is experimentally known that the order-disorder transition temperature of 
heavy ice (D$_2$O) is higher than light ice (H$_2$O) by 4 K \cite{Tajima82, Matsuo1986}.
Table \ref{table:Tc} shows that we already observe the phase transition with calculations at the classical limit of free energy, for both PBE and vdW-DF$^{\rm{PBE}}$
approximations, and that the transition temperature decreases with the inclusion of zero-point effects.
The vdW-DF$^{\rm{PBE}}$ results are below the glass transition temperature  around 100-110 K \cite{Wooldridge1988, Suga1997}
where proton mobility diminishes. This is in general agreement with the experimental order-disorder phase transition temperatures.
Although PBE gives a correct prediction of the stable phase, and an isotope effect of 6.4$\%$,
the value of the phase transition temperature is much larger than the experimental range.
In agreement with the experimental 4 K difference in the phase transition temperature of the isotopes,
the vdW-DF$^{\rm{PBE}}$ functional predicted transition temperature of the heavy ice is larger than the light ice with a 6 K difference.
As a result, with this method, the ratio between the phase transition temperatures of heavy and light ice is reproduced within 1\% of the experimental value
and the isotope effect on the temperature with respect to the H$_2$O transition temperature is calculated to be 6.6$\%$,
as compared to the 5.6$\%$ of the experimental isotope effect.
Therefore, it is important to note that inclusion of non-local van der Waals forces is critical for a reasonable prediction of the transition temperature.

\begin{figure}[!ht]
        \centering
        \includegraphics[clip=true, trim=0mm 0mm 0mm -20mm, width=0.48\textwidth]{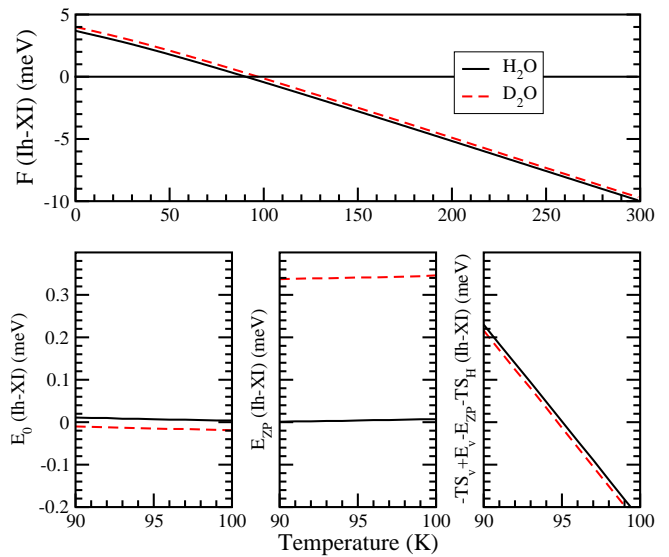}
        \caption{Top, free energy difference per molecule between ice Ih and ice XI calculated with vdW-DF$^{\rm{PBE}}$ functional in the region of the phase transition. Bottom, contributions
to this free energy difference by each term in eq. \ref{eq:free}. Left frozen lattice electronic term. Middle zero-point vibrational energy. Right remaining terms. All the energies on the bottom plots have been shifted to allow them to be compared in the same energy scale.}
        \label{fig:contributionsF_T}
\end{figure}

To understand the main reason behind the difference 
between the transition temperatures of different isotopes,
we study the temperature dependence of 
each component of eq. \ref{eq:free} separately,
as presented in Fig. \ref{fig:contributionsF_T}. 
The electronic energy difference between the two ices, $E_0({\rm Ih})-E_0({\rm XI})$, is larger for H$_2$O than D$_2$O. 
This would result in a larger transition temperature for H$_2$O than D$_2$O, contradicting experiments. 
The last two terms, zero-point-free vibrational entropy and energy $-TS_v +E_{v}-E_{ZP}=\sum_k k_B T \ln \left(1-e^{-\hbar \omega_k(V_{F_{\rm min}}(T)) / k_B T}\right)$, and 
configurational entropy $-TS_H$ also result in larger energy difference for H$_2$O than D$_2$O.
However, the zero-point vibrational energy $E_{ZP}=\sum_k \hbar\omega_k(V_{F_{\rm min}}(T))/2$ difference 
between the two ices, $E_{ZP}({\rm Ih})-E_{ZP}({\rm XI})$, is smaller for H$_2$O than D$_2$O. This is the only term
that shifts the transition temperature of H$_2$O  below that of D$_2$O.
These results show that the phase transition occurs at a lower $T$ for H$_2$O than D$_2$O because of the zero-point energies of the phonon modes.

This can also be seen from a simple model.
The transition occurs when the free energies of the two ices are equal.
For the sake of simplicity, we can set the zero of energy at the frozen lattice cohesive energy of ice XI 
and denoting by $E_d=E_0({\rm Ih})-E_0({\rm XI})$ and $S_H$ the energy and the residual entropy caused by the disorder of ice Ih.
The free energies are $F({\rm XI})=0$ and $F({\rm Ih})=E_d-TS_H$.
Therefore, at the zeroth order, in the classical limit, it follows that $T_c(0)=E_d/S_H$.
When the vibrations are included, the free energies from eq. \ref{eq:free} are equal at the transition temperature,
\begin{equation}
T_c=\frac{E_d}{S_H-k_B \sum_k \ln \left[ \frac{\sinh(\hbar \omega_k({\rm Ih}) / 2 k_B T_c)}{\sinh(\hbar \omega_k({\rm XI}) / 2 k_B T_c)} \right]}.
\end{equation}
A more detailed discussion with the high and low $T$ limits of this transition temperature and the latent heat can be found in the SM \cite{SM}.
Assuming the shifts are not large, the isotope shift in the transition temperature can be simplified to
\begin{equation}
\frac{T_c^{{\rm D_2O}}-T_c^{{\rm H_2O}}}{T_c^{{\rm H_2O}}}=\frac{k_B T_c}{E_d} \sum_k \ln \left[ \frac{R_k ({\rm Ih})}{R_k ({\rm XI}) } \right]
\end{equation}
where
\begin{equation}
R_k ({\rm Ih})=\frac{\sinh(\hbar \omega_k ^{{\rm D_2O}} ({\rm Ih}) / 2 k_B T_c)}{\sinh(\hbar \omega_k ^{{\rm D_2O}} ({\rm Ih}) / 2 k_B T_c)},
\end{equation}
and similarly for $R_k ({\rm XI})$.
Then the low temperature limit becomes,
\begin{eqnarray}
\frac{\Delta T_c ({\rm low})}{T_c^{{\rm H_2O}}}=\frac{\hbar}{2 E_d} \sum_k [ &(&\omega_k ^{{\rm D_2O}} ({\rm Ih}) - \omega_k ^{{\rm D_2O}} ({\rm XI})) \nonumber \\
                                                                         - &(&\omega_k ^{{\rm H_2O}} ({\rm Ih}) - \omega_k ^{{\rm H_2O}} ({\rm XI})) ]
\end{eqnarray}
where the difference in the frequencies exactly corresponds to the energy difference shown in the bottom middle panel of Fig. \ref{fig:contributionsF_T}.
This model clearly shows that the main source of the isotope effect in the transition temperature is the difference in the zero point energies of the different ices.

\begin{figure}[ht]
	\centering
	\includegraphics[clip=true, trim=0mm 2mm 0mm 8mm, scale=0.26]{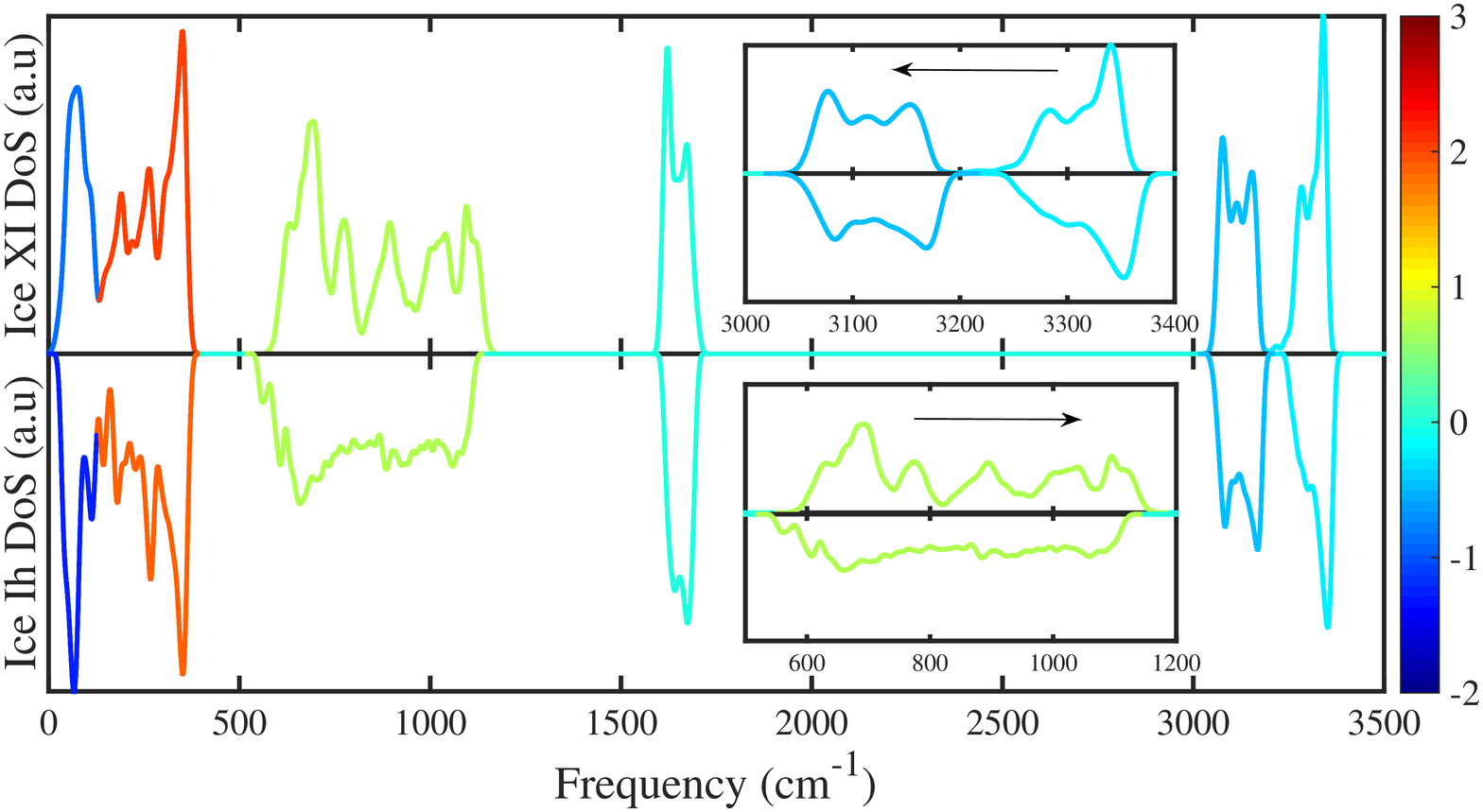} 
	\caption{Vibrational density of states for H$_2$O for proton-ordered ice XI and disordered Ih structures, as obtained with vdW-DF$^{{\rm PBE}}$ functional. 
		Average Gr\"{u}neisen constants of the different modes are given in color code.
		The inset above zooms into the stretching modes and shows the redshift, while
		the inset below zooms into the librational modes and shows the blueshift in ice XI with respect to ice Ih.}
	\label{fig:dos1}
\end{figure}

This is also evident from the phonon density of states of the two ices.
Fig. \ref{fig:dos1} shows phonon density of states for H$_2$O for both
proton-ordered ice XI and proton-disordered ice Ih at zero temperature.
The colors represent the average Gr\"uneisen parameter of each band separately.
The main effect driving the isotope differences is associated to 
the blue shift of the librational band in ice XI with respect to ice Ih,
and a corresponding redshift of the stretching band.
Therefore, with proton ordering, the covalency of the intra-molecular bonds is weakened,
while the inter-molecular hydrogen bonding is strengthened.
This combined with the weights of the Gr\"uneisen parameters results in an overall slightly
larger zero point energy for ice XI than ice Ih.

One of the reasons of the quantitative difference from the experimental results of transition temperature 
can be due to the error in the estimation of residual entropy from disorder in both systems.
Experimentally, it has been shown that at the transition, 
ice XI loses much but not all of the entropy at $T_c$ \cite{Whitworth1996}.
However, it is not clear whether this arises from equilibrium thermal disorder in ice XI, or from failure to complete the phase transition,
leaving some domains of non-equilibrium ice Ih coexisting with ice XI \cite{SugaReview}.
Another reason for the quantitative difference can be the loss of precision of the QHA at larger temperatures,
as the temperature dependence of the phonon vibrations is not taken into account.
This is also the case in the calculated $V_0$ with isotope effects; 
the calculated values deviate from the experimental values at larger temperatures \cite{Pamuk12}.
However, this deviation is not significant at around 100 K, which is the region of interest of this work.

Finally, the exact values depend on the choice of the DFT functional.
While we have shown that the inclusion of vdW interaction in the functional is crucial, 
it should be noted that the local part of the XC functional also changes the structure significantly.
In Ref. \onlinecite{Pamuk12}, it has been shown that vdW-DF functional with the local XC flavor of revPBE
softens the structure such that the anomalous isotope effect on the V(T) is not reproduced at low temperatures.
In addition, Ref. \onlinecite{Galli12} studied the phonon dispersion of ice XI.
While the distribution of the modes are almost identical,
the values of the stretching modes are higher and the librational modes are lower by $\sim 50$ cm$^{-1}$ than those calculated in this work.
Furthermore, Ref. \onlinecite{Santra2013} studied the non-local vdW functionals with different GGA and hybrid functionals for the local XC,
and showed that the cohesive energies depend on the choice of these functionals.
A hybrid functional with exact exchange for the local XC,
with a vdW functional for the non-local correlations could be a good candidate to improve on these results.

All in all, the QHA within DFT with non-local vdW forces, predicts a 6 K temperature difference 
between the isotopes, as compared to the experimental 4 K difference.
This isotope shift is solely due to the nuclear quantum effects from the phonon vibrational energy differences,
and it is predicted without invoking any other effects, such as tunneling.

\section{Conclusion}

In this study, we did a detailed analysis of the phase transition 
between the ferroelectric vs. antiferroelectic-proton-orderered ice XI and disordered ice Ih.
\textit{Ab initio} DFT is necessary to correctly predict the most stable phase of ice as the ferroelectric-ordered ice XI.
The TTM3-F force field model needs improvement for the energy predictions, especially at low temperatures.

By including nuclear quantum effects to the free energy, we have predicted the ferroelectric order to disorder phase transition for hexagonal ices.
The best accuracy requires using the vdW-DF$^{\rm{PBE}}$ functional, with a transition temperature at about 91 K for H$_2$O and 97 K for D$_2$O.
This 6 K temperature difference is mainly due to the difference in the zero point energy of ice with different isotopes, while entropy related terms contribute 
in the opposite direction.
The method is robust to correctly predict and explain the isotope effect on the order/disorder phase transition of hexagonal ice Ih.

This work is supported by DOE Grants No. DE-FG02-09ER16052 and DE-SC0003871 (MVFS) and DE-FG02-08ER46550 (PBA),
and the grant FIS2012-37549-C05 from the Spanish Ministry of Economy and Competitiveness.
Research was carried out in part at the Center for Functional Nanomaterials, Brookhaven National Laboratory, 
which is supported by the U.S. Department of Energy, Office of Basic Energy Sciences, under Contract No. DE-AC02-98CH10886.

\bibliography{PaperBetul}

\end{document}